\documentclass{acm_proc_article-sp}
\usepackage{graphicx}
\usepackage{algorithmic}
\usepackage{algorithm}
\usepackage{ amsmath}

\newcommand{\denselist}{
     \setlength{\itemsep}{0pt}
     \setlength{\parsep}{1.5pt}
     \setlength{\topsep}{1.5pt}
     \setlength{\parskip}{2pt}
     \setlength{\partopsep}{0pt}
     \setlength{\labelwidth}{1em}
     \setlength{\labelsep}{0.5em} }

\newcommand{\bdesc}{\begin{description}\denselist}
\newcommand{\edesc}{\end{description}}

\newcommand{\blist}{\begin{itemize}\denselist}
\newcommand{\elist}{\end{itemize}}
\long\def\remove#1{}
\newtheorem{theorem}{Theorem}
\begin{document}
\title{Predicting Influential Users in Online Social Networks}

\numberofauthors{2}

\author{
\alignauthor Rumi Ghosh\\
       \affaddr{USC Information Sciences Institute}\\
       \affaddr{4676 Admiralty Way}\\
       \affaddr{Marina del Rey, CA 90292}\\
       \email{rumig@usc.edu}
\alignauthor
Kristina Lerman\\
       \affaddr{USC Information Sciences Institute}\\
       \affaddr{4676 Admiralty Way}\\
       \affaddr{Marina del Rey, CA 90292}\\
       \email{lerman@isi.edu}
}
\date{}

\maketitle

\begin{abstract}
Who are the influential people in an online social network?  The answer to this question depends not only on the structure of the network, but also on details of the dynamic processes occurring on it. We classify these processes as \emph{conservative} and \emph{non-conservative}. A random walk on a network is an example of a conservative dynamic process, while information spread is non-conservative. The influence models used to rank network nodes can be similarly classified, depending on the dynamic process they implicitly emulate. We claim that in order to correctly rank network nodes, the influence model has to match the details of the dynamic process.
We study a real-world network on the social news aggregator Digg, which allows users to post and vote for news stories. We empirically define influence as the number of in-network votes a user's post generates. This influence measure, and the resulting ranking, arises entirely from the dynamics of voting on Digg, which represents non-conservative information flow.
 We then compare predictions of different influence models with this empirical estimate of influence. The results show that  {non-conservative}  models are better able to predict influential users on Digg.
We find that \emph{(normalized)} $\alpha$-\emph{centrality} metric turns out to be one of the best predictors of influence. We also present a simple algorithm for computing this metric and the associated mathematical formulation and analytical proofs.
\end{abstract}

\section{Introduction}
Online social networks have become important hubs of social activity and conduits of information.
Popular social networking sites such as Facebook, the social news aggregator Digg, and the microblogging service Twitter have undergone explosive growth.
Though some research suggests that people are more affected by the opinions of their peers than influentials~\cite{Domingos01, Watts:2007}, recent studies of online social networks~\cite{Cha:2010} support the hypothesis that influentials exert disproportionate amount of influence.
With the numbers of active users on these sites numbering in the millions or even tens of millions, identifying influential users among them becomes an important problem with applications in marketing~\cite{Kempe03}, information dissemination~\cite{Gruhl04, Leskovec07kdd}, search~\cite{Adamic05search}, and expertise discovery~\cite{Davitz07}.

\remove{
A \emph{dynamic process} is a sequence of continuously changing conscious or unconscious acts of collaborations between interdependent participating entities within a network(system), jointly shaping the intermittent and net outcome of these actions. Each independent participating entity may have some initial mass, which might change as an outcome of the entity's participation in the dynamic process. This change of mass is due to the  flow of mass or traffic \cite{Borgatti:2005} between the interconnected entities during a dynamic process.
}

While many influence models and centrality measures have been proposed to rank actors within a social network, almost all of them make implicit assumptions about the underlying dynamic process occurring on the network~\cite{Borgatti:2005}, which may not be applicable to online social networks. 
\remove{Take, as an example, PageRank~\cite{PageRank}, a popular metric used for finding influential nodes in a network, e.g., pages on a Web graph. This metric is known to describe a random walk~\cite{?}. While the behavior of a Web surfer may be modeled as a random walk, other dynamic processes, such as information dissemination, may not be.}
Since dynamic processes can often be directly observed on online social networks~\cite{Lerman10icwsm,Lee10}, these networks provide us with a unique opportunity to study influence. In this paper we address the question of which influence model is most suitable to predict the influence standings of users within online social networks whose main function is to disseminate information.

We classify {dynamic processes}, or flows, that can occur on social networks as  \emph{conservative} or \emph{non-conservative}. We define a flow to be {conservative} (``transfer'' in Borgatti's ~\cite{Borgatti:2005} terminology) if the initial mass or content of the network
 is equal to the final mass after the flow has taken place. For example, consider that each user $u_{i}$ in a social network has some amount of money $m_{i}$, some fraction of which she can transfer to any of her friends.  The total amount of money in the network remains constant($\sum m_{i} =c$) at all times.  Hence, the money exchange process within a social network is {conservative}. We define a flow to be {non-conservative} (``parallel'' or ``serial'' duplication in Borgatti's terminology) if the initial mass of
the system  is not equal to the final mass after the flow has taken place. Information flow is a  {non-conservative} process. When a user posts a new item on Facebook, Digg or Twitter, she broadcasts this information to her social network. Each of her social links may in turn broadcast the information to their own social networks, thereby continuing the parallel duplication process. Note that this process is somewhat different from gossip, which is one-to-one, or serial duplication of information. Both types of information spread in online social networks, however, are {non-conservative} processes.


Influence models  too, can be categorized as {conservative} and {non-conservative}. Though these models take only network structure into account when measuring importance or centrality of an actor within the network, they make assumptions about the details of the underlying dynamic process taking place on the network. For example, PageRank algorithm~\cite{Brin:1998}, commonly used for network analysis, models a conservative diffusion process; therefore, it may not be an appropriate ranking metric for online social networks. In order to correctly rank influential nodes in a network, the influence model has to match the details of the dynamic process.

This paper makes three contributions.
In Section~\ref{sec:influence} we review many of the existing influence models found in literature and categorize them according to whether they model a conservative or a non-conservative dynamical process.
In Section~\ref{sec:methodology} we study a real-world network on the social news aggregator Digg. Digg allows users to post and vote for news stories, and also to create networks in order to track what new stories their friends posted or voted for. We define \remove{a statistically significant} an empirical measure of influence as the number of in-network votes a user's post generates, i.e., votes that come from  that user's social network links. This influence measure, and the resulting ranking, arises entirely from the dynamics of voting on Digg.
In Section~\ref{sec:comparison} we evaluate the different influence models by correlating the rankings they produce with the rankings produced by the empirical measure of influence.
We find that the non-conservative  
$\alpha$-\emph{centrality}~\cite{Bonacich:2001} best predicts the rankings of Digg users.
These results corroborate our claim that the details of the influence model used for ranking actors in a network should match the details of the dynamic process occurring on the network. We review related works pertaining to each section in the section itself.

There exist many empirical studies of social behavior and influence on online social networks. Some of these studies, compare empirical measures of influence with some structural models of influence like PageRank or in-degree centrality~\cite{Lee10,Cha:2010}.
However, there is a need to  clearly differentiate between the two distinct and different methods of quantifying influence in online social networks:
\vspace{-10 pt}
\begin{enumerate}
\item Measurements of online social behavior or the dynamic processes occurring on a social network  to estimate influence.
\item Using influence models based on the structural properties of the underlying social network to predict influence.
\end{enumerate}
\vspace{-10 pt}
\noindent Empirical estimates of influence measured from online social behavior, do not have the predictive capabilities of the structural models of influence.
 To the best of our knowledge,  ours is the first work that evaluates predictive influence models based solely on structural properties of the underlying social network,  using the actual dynamic process occurring on a real-world network;  unlike existing works, which simulate the underlying dynamic process~\cite{Borgatti:2005,Kiss:2008}.

We also provide a simple method to calculate the \emph{(normalized)} $\alpha$-\emph{centrality} and provide a mathematical validation and analytical proofs for this method.

\section{Influence Models}
\label{sec:influence}
A network  of $n$ actors and $m$ links can be represented as a graph $G(V,E)$ of $V (|V|=n)$ nodes and $E (|E|=m)$ edges. Each actor is represented as a node. An edge exists between node $i$ and $j$ if actor $i$ is linked to actor $j$. The edges might be weighted to exhibit the strength of the links.  The geodesic distance from $i$ to $j$ is $gd(i,j)$. Let $d^{in}_{i}$ be the in-degree of node $i$. Let $d^{out}_{i}$ be the out-degree of node $i$. If there exists a directed edge $e_{ij}$ from $i$ to $j$, we say that $i$ is a fan of $j$ and $j$ is a friend of $i$. Let $A=(A_{ij})$ be the adjacency matrix of the corresponding network, whose maximum eigenvalue is $\lambda_{1}$ and whose maximum out- and in-degrees are $d^{out}_{max}$ and $d^{in}_{max}$ respectively.

Below we review existing influence models found in literature and categorize them as conservative or non-conservative according to the flow that they model.

\subsection{Geodesic Path-based Ranking Measures}
All geodesic path-based ranking methods assume that network flow is  \emph{conservative} in nature. Moreover, these methods assume a binary flow. We define a binary dynamical process, D, as follows:
\vspace{-10 pt}
\blist
\item There exists some initial mass at some node $i$ ($m_{i}=M_D$) at time $t_{0}$.
\item At any time, she may either transfer the entire mass to a single neighbor $k$ ($m_{k}=M_{D}$) or keep it to herself ($m_{j}=M_{D}$).
\item The length of a path traversed while moving from node $i$ to node $j$ via edge $e_{ij}$ $ \forall e_{ij} \neq 0$ is equal to the weight of the edge from node $i$ to node $j$.
\elist
\vspace{-20 pt}
\noindent \paragraph{Closeness Centrality}
Let node $i$ generate a sequence of binary processes $(D_{i j_1},D_{ij_2},\cdots,D_{ij_p},\cdots)$,  with the objective of process $D_{ij_p}$ being to transfer the initial mass $M_{D_{ij_p}}$ at node $i$, to another node $j_{p}\ (j_{p}\neq i)$. For binary flow  process $D_{ij_1}$, with the initial mass at node $i$ ($m_{i}=M_{D_{ij_1}}, m_{k}=0, \forall k\neq i $),  the geodesic distance $gd(i,j_{1})$ is the shortest  distance traversed
by the mass in reaching destination node $j_{1}$.  When this  mass reaches $j_{1}$, node $i$ generates another flow, $D_{ij_2}$ ($m_{i}=M_{D_{ij_2}}, m_{j_{1}}=M_{D_{ij_1}}, m_{k}=0\  \forall k\neq i,j_{1} $), with the objective that the initial mass $M_{D_{ij_2}}$, be  transferred to another node $j_{2}\ (j_{1}\neq j_{2}, i)$, which does not have any mass. Continuing with this sequence of binary flows, when mass $M_{D_{i,j_{p-1}}}$ reaches node $j_{p-1}$, node $i$ generates another flow, $D_{ij_{p}}$ ($m_{i}=M_{D_{ij_{p}}}, m_{j_{k}}=M_{D_{i j_{k}}} \forall {j_k}\neq i, k<p, m_{j_k}=0  \forall {j_k}\neq i, k \ge p $), with the objective that the initial mass $M_{D_{ij_{p}}}$, be  transferred to another node $j_{p}\ (j_{p}\neq i, j_{k}\  \forall k<p)$, which does not have any mass. This process is terminated when every  node connected to node $i$ has some mass. Closeness centrality of node $i$ is inversely proportional to the shortest total distance traversed by all masses transferred from node $i$ to some connected node $j$, when this sequence of binary processes terminate.

There are different definitions of closeness centrality in literature. Hakimi~\cite{Hakimi:1965} and Sabidussi~\cite{sabidussi:1966} defined \emph{closeness} centrality as
\begin{equation}
C_{c}(i)=\frac{1}{\sum_{j=1}^{n} gd(i,j)}
\end{equation}
\noindent In order to discount network size, Wasserman and Faust~\cite{Wasserman:1994} modified the definition of closeness centrality to
\begin{equation}
C_{c}(i)=\frac{n-1}{\sum_{j=1}^{n} gd(i,j)}
\end{equation}
\noindent These closeness centrality measures implicitly assume that the underlying social network is strongly connected. However, this assumption does not hold most real-life network. Therefore, Lin~\cite{Lin:1976} redefined closeness centrality using  the number of nodes reachable from node $i$, $J_{i}$:
\begin{equation}
C_{c}(i)=\frac{\frac{J_{i}}{n-1}}{\frac{\sum_{j=1}^{n} gd(i,j)}{J_{i}}}
\label{lin}
\end{equation}
\vspace{-10 pt}
\noindent \paragraph{Graph Centrality}
Let node $i$ generate a binary flow $D_{i}$ with the objective to transfer mass $M_{D_{i}}$ to a node $j$ at largest geodesic distance from it. Graph centrality is inversely proportional to the distance traversed by mass $M_{D_{i}}$, to move from node $i$ to node $j$.
Formally,  graph centrality is defined as \cite{hage:1995}:
\begin{equation}
C_{g}(i)=\frac{1}{\max_{j \in V\setminus \{i\}} gd(i,j)}
\end{equation}
\vspace{-10 pt}
\noindent {\paragraph{Betweenness Centrality}
Let every node $j$ generate a  sequence binary flows $(D_{jk})$ with the objective to transfer mass $M_{D_{jk}}$ to node $k\ (k \in V\setminus \{j\})$. Let each of these processes take the shortest route, i.e., the path of transfer of mass $M_{D_{jk}}$ from $j$ to $k$ is the geodesic path $gd(j,k)$. Betweenness centrality of node $i$ is proportional to the total number of times a given node $i$ is traversed by all these processes (excluding processes that start or end at $i$).
Formally,  betweenness centrality is defined as \cite{freeman:1977a}
\begin{equation}
C_{b}(i)=\sum_{i \neq j \neq k}\frac{\sigma_{jk}(i)}{\sigma_{jk}}
\end{equation}
where $\sigma_{jk}$ is the number of geodesic paths from $j$ to $k$ and $\sigma_{jk} (i)$ are the number of shortest paths from $j$ to $k$ which traverse $i$.}

\subsection{Topological Ranking Measures}
The topology of a network  is characterized by the  inherent structural properties of the nodes and edges comprising it. In case of a node,  it includes the node's  in- and out-degree. In case of an edge $e_{ij}$,  it includes the out-degree of node $i$ and in-degree of node $j$. The geodesic path-based ranking measures do not take the topology of the network into account. This is due to the binary nature of the underlying process, i.e., either the entire mass is transferred from node $i$ to a single neighbor $j$ or it is not transferred at all. This transfer is independent of the number of neighbors of $i$ and $j$. However, there do exist topological ranking measures, as described below.

\subsubsection{Markov Process-based Ranking Measures}
Markov processes describe a broad class of random processes, including random walks and diffusion processes.
In a Markov process, the probability of transfer from node $i$ to $j$ depends only on the state of $i$, which is described by the topology of node $i$.

Let $P=(P_{ij}) $ define the transition matrix of a Markov process.
Let  $M_{D_{i}}$ be  the number of  Markov processes generated at node $i$.
Let $m_{ij}^{t_{k}}$ be the mass in node $j$ from process $D_{i}$ (generated by node $i$) at time $t_{k}$. Consider the following \emph{conservative} flow $D_{i}$:
\vspace{-10 pt}
\blist
\item The initial mass is at node $i$, $m^{t_{0}}_{ii}=M_{D_{i}}$ at time $t_{0}$.
\item At time $t_{1}$, $i$ transfers a fraction $P_{ij}$ of mass $M_{D_{i}}$ to her friend $j$ ($m^{t_1}_{ij}=P_{ij}M_{D_{i}}$). She may also retain some  fraction of mass $\sum_{k=i, j \in \mbox{friend}( i)} P_{ik}=1$.
\item Similarly, at any  time $t_{k}$, node $j$ transfers some fraction $P_{jp}$ of its mass  to her friend ($p \in \mbox{friend}(j)$) and keeps some fraction of the mass to herself.
\elist

This flow conserves the total mass in the system at any given time, $\sum_{j \in V} m^{t_{k}}_{ij}=M_{D_{i}}$.
The expected number of  Markov processes generated at $i$ reaching node $j$ at time $t_k$ is equal to $m^{t_{k}}_{ij}$. Hence the flow underlying a Markov process is \emph{conservative}.

\paragraph{PageRank}
Let
\begin{eqnarray}
P_{ij} & =& \alpha/ d^{out}_{i}  \mbox{ } \forall j\neq i \mbox{ \& } A_{ij}=1 \nonumber \\
&= & 1-\alpha  \mbox{ if } i=j \nonumber \\
& = & 0\mbox{ otherwise}
\end{eqnarray}
\noindent Let each node $i$ generate a flow $D_{i}$. Each flow terminates at time $t_{k_{i}}$ such that  $ m_{ij}=m^{t_{k_{i}}}_{ij}=m^{t_{k_{i}-1}}_{ij}\forall i, j\in V$. Let $m_{j}= \sum_{i\in V} m_{ij}$ be the mass at node $j$ when all these processes terminate where $ \sum_{j\in V}m_{j}=\sum_{j\in V} M_{D_{j}}$. $\alpha$ is called the \emph{damping factor}.  PageRank ranks node $i$ proportional to the  mass $m_{i}$  at node $i$ at the end of these processes \cite{Brin:1998}.
\begin{eqnarray}
C_{pr, \alpha}(i) & = &(1-\alpha)+\alpha \sum_{j \in fan(i)} \frac{C_{pr, \alpha}(j)}{d^{out}_{j}}  \nonumber \\
C_{pr, \alpha} &= &(1-\alpha)e+\alpha  C_{pr, \alpha} P
\label{eq:pr}
\end{eqnarray}
where $C_{pr, \alpha}$ is the $(1\times n)$ \emph{PageRank vector},  $e$ is the $(1\times n)$ \emph{unit} vector and $P$ is the $(n\times n)$ \emph{transition matrix}.
Using a \emph{personalization} vector $v$ \cite{Brin:1998a} instead of $e$, we have personalized PageRank as
\begin{equation}
C_{pr, \alpha} = (1-\alpha)v+\alpha  C_{pr, \alpha} P
\label{eq:ppr}
\end{equation}

\paragraph{Hubbels Model}
A very similar model was proposed by Hubbel in 1965~\cite{Hubbell:1965}:\\
\begin{equation}
C_{h}^{T} = v^{T}+  PC_{h}^{T}
\end{equation}
where $C_{h}$ is Hubbel's ranking vector.

\subsubsection{Degree Centrality}
Consider a \emph{non-conservative} flow  $D_{i}$ as follows:
\vspace{-10 pt}
\blist
\item The process consists of a single time interval $[t_{0}, t_{1}]$.
\item  If the process is generated at node $i$ who has mass $m_{i}=M_{D_{i}}$, at time $t_{0}$, this mass is duplicated to all her fans at time $t_{1}$. Therefore the total mass of the system at time $t_{1}=d^{in}_{i}M_{D_{i}}$ which is proportional to in-degree centrality of $i$.
\elist
\vspace{-10 pt}
\noindent Let each node $i \in V$ start a flow. The in-degree centrality of node $i$ is
$C_{d^{in}}(i) =d^{in}_{i}$.

Similarly if we consider another \emph{non-conservative} flow $D_{j}$ as follows:
\vspace{-10 pt}
\blist
\item The process consists of a single time interval $[t_{0}, t_{1}]$.
\item  If the process is generated at node $j$ who has mass $m_{j}=M_{D_{j}}$, at time $t_{0}$, this mass is duplicated to all her friends at time $t_{1}$. Therefore the total mass of the system at time $t_{1}=d^{out}_{j}M_{D_{j}}$ which is proportional to out-degree centrality $j$.
\elist
\vspace{-10 pt}
Let each node $i \in V$ start a flow. The out-degree centrality of node $i$ is
$C_{d^{out}}(j) =d^{out}_{j}$.

\subsubsection{Path-based Ranking Measures}
Consider a \emph{non-conservative} flow $D_{i}$ defined as follows:
\vspace{-10 pt}
\blist
\item $D_{i}$ is generated by node $i$ at time $t_{0}$, $m^{t_0}_{ii}=M_{D_{i}}$.
\item At time $t_{1}$, let $\alpha^{t_{1}}_{ij}$ be the fraction of mass at $i$ duplicated by $j$ where $j$ is a friend of $i$.   Therefore the mass  at $j$ at time $t_{1}$  due to process $D_{i}$ is  $m^{t_1}_{ij}$ where $m^{t_1}_{ij}= \alpha^{t_{1}}_{ij} \cdot M_{D_{i}}$.
\item Similarly let $\alpha^{t_{k}}_{qj}$ be the fraction of mass at $q$ duplicated by friend $j$, at time $t_{k}$. Therefore the mass  at $j$ at time $t_{k}$ is  $\sum^{k}_{p=0}m^{t_p}_{ij}$ where $m^{t_p}_{ij}= \sum_{q \in fans(j)} \alpha^{t_{p}}_{qj}m^{t_{p-1}}_{iq}$,\ $ p>0$.
\elist
\vspace{-20 pt}

\paragraph{$\alpha$-Centrality}
Let each node $i$ generate a flow $D_{i}$ described above. If these {non-conservative} flows persist for a long time with with $M_{D_{i}} =v_{i}$ and $\alpha^{t_{p}}_{ij} =\alpha \forall p$, then the mass at $i$, $m^{t_{k} \to \infty}_{i}$ is proportional to the rankings given by $\alpha$-centrality. Formally, $\alpha$-centrality defined by Bonacich~\cite{Bonacich:2001} is:
\begin{equation}
C_{alpha, \alpha} = v+\alpha  C_{alpha , \alpha}  A =v{(I-\alpha A)}^{-1}
\label{a-cen}
\end{equation}
\noindent where $C_{alpha, \alpha}$ is the $\alpha$-centrality vector.
$\alpha$-centrality can also be written as:
\begin{eqnarray}
C_{alpha, \alpha} & = & v(\sum^{k\to \infty}_{t=0} \alpha^{t} A^{t}) \\
&=& vC_{\alpha, k \to \infty}
\label{eq:b-cen}
\end{eqnarray}
\noindent where $C_{\alpha, k}=\sum^{k}_{t=0} \alpha^{t} A^{t}$ is the $\alpha$-centrality matrix. Hence, for equation \ref{a-cen} to hold, the series $\{C_{\alpha, k}\}$  must necessarily converge as $k$ approaches infinity. This happens if and only if  $|\alpha| < \frac{1}{|\lambda_{1}|}$. Therefore $\alpha$-centrality can only be calculated for  $|\alpha| < \frac{1}{|\lambda_{1}|}$~\cite{Bonacich:2001}. Bonacich states that $\alpha$ ``reflects the relative importance of endogenous versus exogenous factors in the determination of centrality.'' Following Katz~\cite{Katz:1953}, we call $\alpha$ the \emph{attenuation factor}.

\paragraph{Normalized $\alpha$-centrality}
In this paper we define \emph{normalized\ $\alpha$-centrality} as:
\begin{equation}
C_{N_{\alpha}, \alpha} =\frac {vC_{ \alpha,k \to \infty}}{\sum^{n}_{i,j} {(C_{\alpha, k \to \infty})}_{ij} }
\label{eq:norm-cen}
\end{equation}

As stated above, the value of $\alpha$ is subject to the constraint $|\alpha| < \frac{1}{|\lambda_{1}|}$.
We show that computation of normalized $\alpha$-centrality is not bounded by this constraint (Section~\ref{sec:cal_influence}). However, rankings given by normalized $\alpha$-centrality are equal to the rankings given by $\alpha$-centrality for $|\alpha| < \frac{1}{|\lambda_{1}|}$. We also show that  value of  normalized $\alpha$-centrality remains the same $\forall \alpha \in (\frac{1}{|\lambda_{1}|},1] (C_{N_{\alpha},\alpha > \frac{1}{|\lambda_{1}|}}=C_{N_{\alpha}}$) and is independent of $\alpha$. We further show that when $|\lambda_{1}|$ is strictly greater than any eigenvalue, $\lim_{\alpha  \to \frac{1}{|\lambda_{1}|}}C_{N_{\alpha}, \alpha}$ exists and $\lim_{\alpha  \to \frac{1}{|\lambda_{1}|}}C_{N_{\alpha}, \alpha} =C_{N_{\alpha}}=C_{N_{\alpha}, \alpha >\frac{1}{\left |\lambda_{1}\right | }  } $.
The details of the mathematical formulation of this metric and the associated proofs are given in the Appendix.

\paragraph{Katz Score}
If $v=\alpha eA$, $\alpha$-centrality reduces to Katz score \cite{Katz:1953}.
\begin{equation}
C_{k, \alpha} = \alpha eA{(I-\alpha A)}^{-1}
\label{Katz}
\end{equation}

\paragraph{SenderRank}
SenderRank~\cite{Kiss:2008} has a similar flavor being defined as:
\begin{equation}
C_{SR, \alpha} ^{T}=(1-\alpha){(I-\alpha A)}^{-1}e^{T}
\label{eq:sr}
\end{equation}
\noindent where $C_{SR, \alpha}$ is the SenderRank vector.

\paragraph {EigenVector Centrality}
EigenVector Centrality~\cite{Bonacich:1972} is given by:
\begin{eqnarray}
C_{E}(i) &= & \frac{1}{\lambda_{1}} \sum^{n}_{j=1}C_{E}(j)A_{ji} \nonumber \\
C_{E}&=&\frac{1}{\lambda_{1}} C_{E}A
\end{eqnarray}
\noindent where $C_{E}$ is the eigenvector centrality vector since it is equal to  an eigenvector of $A$ (corresponding to $\lambda_1$). Most real-life networks such as online social networks have asymmetric relations. Bonacich~\cite{Bonacich:2001} showed that the eigen-vector centrality approach does not work well for asymmetric relations.

\section{Empirical Estimate of Influence}
\label{sec:methodology}
We study a real-world online social network on Digg. Digg is a social news aggregator that enables users to collectively moderate stories they find online by submitting them to Digg and voting for them. Digg promotes the best stories (i.e., stories that receive many votes) to its front page. In addition, Digg allows users to create social networks by adding as friends, users whose activities they want to track. Using the Friends Interface, a user can see the stories her friends recently submitted or voted for.

We rank Digg users according to the centrality measures defined above and compare these rankings with those produced by the empirical estimate of influence.  Our objective is to find the influence model that best predicts influential users in this network.

\subsection{Data Collection}
\label{collection}
We used Digg API to collect data about 3,553 stories promoted to the front page in June 2009. The data associated with each story contained story title, story id, link, submitter's name, submission time, list of voters and the time of each vote, the time the story was promoted to the front page. In addition, we collected the list of voters' friends. From this information, we were able to reconstruct the network of Digg users who were active during the sample period. Borrowing the concept of active user from media research \cite{Levy:1985}; we define an \emph{active user} as a person who votes in at least one story. Next, we get the connections between the active users. We say user $a$ is $connected$ to user $b$ if he is either a \emph{friend} or \emph{fan} of user $b$.  We store an active user in \emph{active users network} if he is connected to one or more active users. In our dataset, there are 139,410 distinct voters who have voted on at least one story. Out of these users, 69,524 voters are $connected$ to one or more active users and hence are members of the \emph{active users network}. These 69,524 $connected$ users form the underlying \emph{friendship network}. Of these 57,908 users form one giant connected component. The diameter  of the network (length of the longest shortest path) is 16. Thus we observe empirically, that the network exhibits \emph{small world phenomena} ~\cite{Milgram:1967} ~\cite{Watts:1999} since the diameter of the network is $O(log\mbox{ }n)$ (n=69,524). $572$  of the $587$  distinct submitters belong to this friendship network.

\subsection{Estimation of Influence}
\label{sec:empirical_influence}

When a user (submitter) posts a story on Digg her fans are able to see the story.
Some of these fans will like the story and vote for it. The story will then become visible to their own fans, who may themselves choose to vote for the story, and so on. Therefore, the underlying dynamic process of information spread on the network is non-conservative in nature.

Assume that user $i$ posts a story.
If a link $e_{ji}$ exists from $j$ to $i$,  then user $j$ is a fan of user $i$ and is watching her activities. In other words, when $i$ posts a story, $j$ is able to see it through Digg's Friends Interface. If $j$ also votes for the story, we call her vote a \emph{fan vote}.
The probability that a submitter's fan votes on a story depends on
\vspace{-10 pt}
\begin{enumerate}
\item  the influence of the submitter
\item the quality of the story
\end{enumerate}
\vspace{-10 pt}
\noindent We assume that story quality is a random variable, uncorrelated with the submitter. Therefore, we can average out the contribution of story quality to submitter's influence, by aggregating fan votes over all stories submitted by the same user.

Let $N$ be the total number of users in the network  ($N=69,524$)
and $K$ be the number of fans the submitter $i$ of story $s_i$ has.  Let $k$ be the total number of  fan votes that story $s_i$ receives within the first $n$ votes. We set $n=100$, calculating the number of fan votes within the first 100 votes.  The stochastic process of voting is described by the \emph{urn model}, in which $n$ balls are drawn without replacement from an urn containing $N$ balls in total, of which only $K$ balls are white. The probability that $k$ of the first $n$ votes are from submitter's fans purely by chance is equivalent to the probability that $k$ of the $n$ balls drawn from the urn are white.
Hence, the probability that $X=k$ of the first $n$ votes are from submitter's fans, $P(X=k|K,N,n)$, is given by the hypergeometric distribution:
 \begin{equation}
P(X=k| K,N,n)= \frac{ \left( \begin{array}{c}
K  \\
k  \end{array} \right)  \left( \begin{array}{c}
N-K  \\
n-k  \end{array} \right)}{\left( \begin{array}{c}
N  \\
n  \end{array} \right)}
\label{eq:hypergeometric}
\end{equation}

 \begin{figure}[tbh]
\begin{tabular}{c}
  \includegraphics[height=1.8 in, width= 3.2 in]{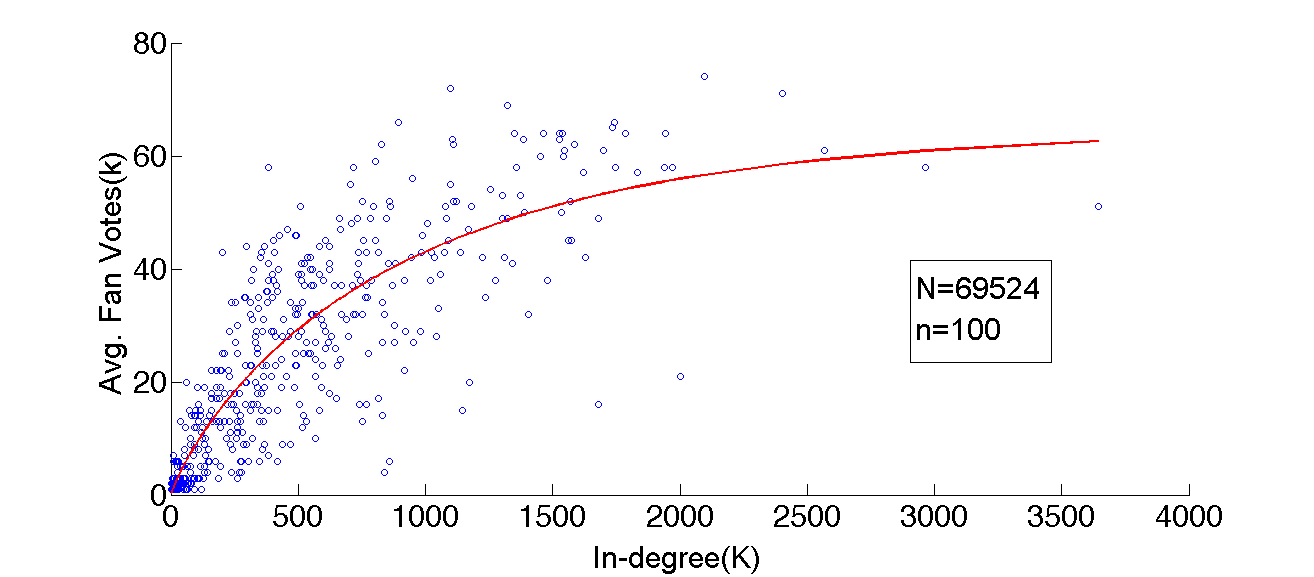}\\
  (a)\\
   \includegraphics[height=1.8 in, width=3.2 in]{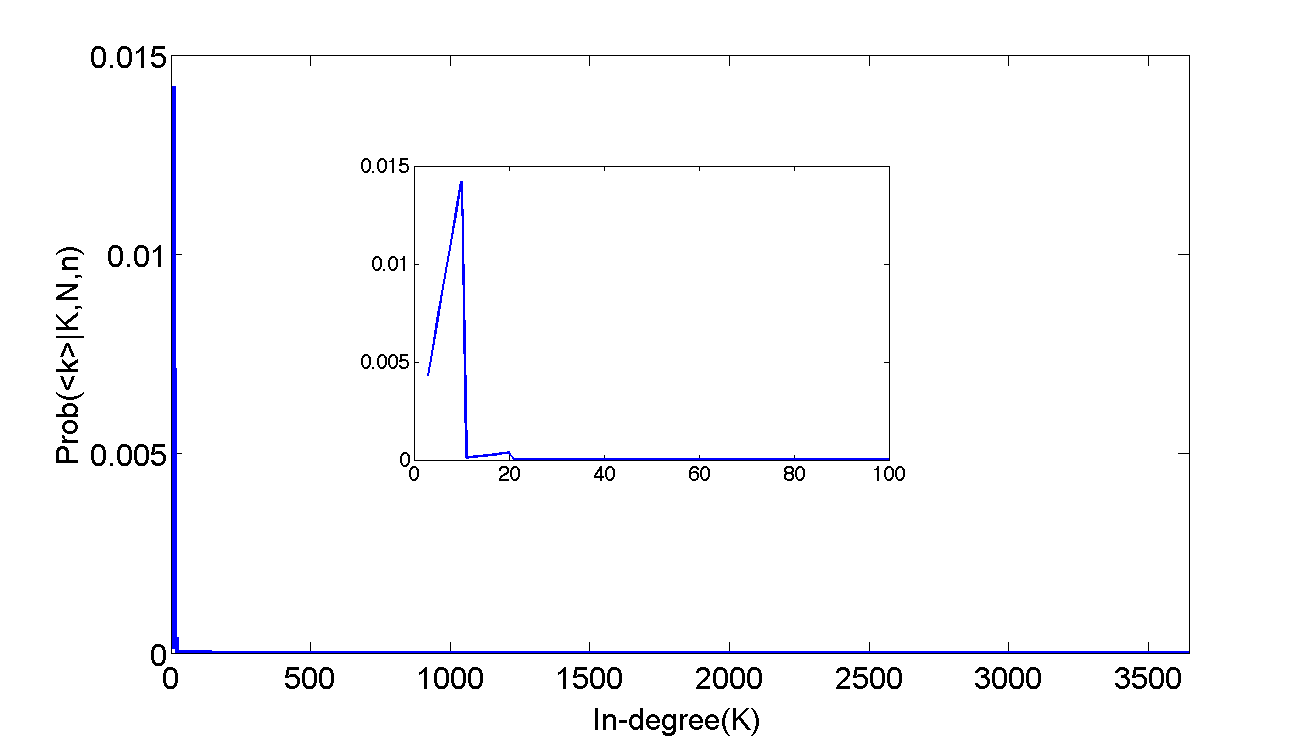}  \\
   (b)\\
\end{tabular}
\caption{(a) The scatter plot shows the average number of fan votes received by a story within the first 100 votes vs submitter's in-degree (number of fans). Each point represents a distinct submitter. The line gives the expected number of fan votes given the in-degree, which can be approximated ($r^2=0.75$) by a Weibull cumulative distribution. (b) This plot shows the probability of the expected number of fan votes being generated purely by chance. The inset zooms in for in-degree less than 100.}\label{fig:avg_fan_votes}
\end{figure}

\remove{
\begin{figure}[tbh]
 \includegraphics[height=2.8in]{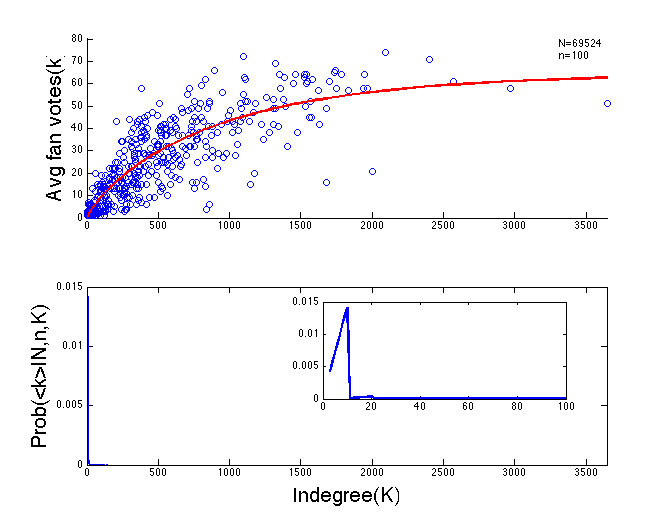}
\caption{ The top scatter plot shows the average number of fan votes received by a story within the first 100 votes vs submitter's in-degree (number of fans). Each point represents a distinct submitter. The line gives the expected number of fan votes given the in-degree, which can be approximated ($r^2=0.75$) by a Weibull cumulative distribution. The bottom plot shows the probability of the expected number of fan votes being generated purely by chance. The inset zooms in for in-degree less than 100.}\label{fig:avg_fan_votes}
\end{figure}
}

Of 3552 stories in our data set,  3489 were submitted by 572 users within the social network.  Of these, 504 submitters received at least one fan vote in the first 100 votes for one of their stories. These 504 users submitted a total of 3396 stories.
For each of the 504  distinct submitter having fan votes in top 100 votes, we plot the average number of fan votes received by stories submitted by these 504 users vs user's in-degree in Figure \ref{fig:avg_fan_votes}.  This scatter plot is approximated by the Weibull cumulative distribution  ($r^2 = 0.75$),
$\langle k \rangle =65(1-e ^{-{(0.0011K+0.0005)}^{0.86}})$.
We use this expression to estimate the expected number of fan votes $\langle k \rangle$ within the first 100 votes for a user with in-degree ($K$) (N=69524, n=100).
Using \ref{eq:hypergeometric}, we then calculate the probability that $\langle k \rangle$ fans voted purely by chance. As can be clearly seen in the lower plot in  Figure~\ref{fig:avg_fan_votes}, for $K>10$, the probability that $\langle k \rangle$ of submitter's $K$ fans voted purely by chance is exceedingly small ($P<0.00038$), and therefore, highly unlikely. We conclude that averaging fan votes over all stories submitted by a given user is an effective indicator of her influence (given she has at least 10 fans).

In our dataset, in order to mask the effect of story quality, we consider only those users who submitted at least two stories. There were 289 distinct submitters with more than two stories which received at least one fan vote within the first 100 votes. All these submitters had more than 10 fans. We use the average number of fan votes that stories submitted by these users received (within the first 100 votes) as an indicator of their influence. We then rank submitters according to this empirical measure of influence.

Influence can be similarly measured in other social networks. For instance, in \cite{Cha:2010, Lee10}, influence in Twitter is measured for the dynamic process of information propagation using retweets and mentions. However, these studies have not proven the statistical significance of the measures employed.

\section{Comparison of Influence \\Models}
\label{sec:comparison}

In most situations, data detailing the history of a dynamical process on a network is not available; therefore, calculating an empirical estimate of influence is not feasible. Instead, many approaches were developed to identify important or influential actors solely using the structure of the network.   Some attempts have been made to analyze these approaches by simulating the underlying dynamical processes \cite{Borgatti:2005,Kiss:2008}. We on the other hand, evaluate these influence models by comparing them with the empirical measure of influence obtained from the analysis of the actual dynamic process of information diffusion on an online social network. Since information propagation is a \emph{non-coservative} process, we hypothesize that \emph{non-conservative} models will best predict the influentials within the network.

\subsection{Calculating Influence}
\label{sec:cal_influence}
It can be easily shown that $\alpha$-centrality is a generalization of Katz score (equation \ref{Katz}).
The computation of $\alpha$-centrality is restricted to $\alpha < \frac{1}{|\lambda_1|}$ \cite{Bonacich:2001}.
However, computing an eigenvalue $\lambda_1$, of the network adjacency matrix $A$ is a costly and time consuming process. Besides, the value of $ \frac{1}{|\lambda_{1}|}$ turns out to be very small in most large networks. In this paper, present a simple algorithm to calculate the normalized $\alpha$-centrality, which is not bounded by this  tight constraint. However, for a given value of $\alpha < \frac{1}{|\lambda_1|}$, the rankings given by normalized $\alpha$-centrality are equal to the rankings given by $\alpha$-centrality (Section \ref{Appendix}, theorem \ref{th1}). As $\alpha$ is increased ($\alpha > \frac{1}{|\lambda_1|}$), $C_{N_{\alpha}, \alpha}$ converges to $C_{N_{\alpha}}$ which is independent of $\alpha$ (Section \ref{Appendix}, theorem \ref{th2}).
A simple algorithm for computing  normalized $\alpha$-centrality as $\alpha$ is varied, using dynamic programming, is given below.
\begin{algorithm}
\caption{ Normalized $\alpha$-centrality}
\label{alg:1}
\begin{algorithmic}
\STATE{\bf{Input}}\\
$A$: Adjacency matrix  \\
$v$: Personalization vector  \\
$s$: Step size (can be modified depending on the granularity of the results desired) \\
$k$: Maximum number of iterations in each step (Equation \ref{eq:b-cen}).\\
$\epsilon$: Tolerance
\STATE{\bf{Output}}\\
$\{C_{N_{\alpha}, \alpha_{t}} : \alpha_{t} \in [0,1] \}$
\STATE  {\bf{Initialize }}\\
 $C_{N_{\alpha},\alpha_{0}}, C^{0}_{N_{\alpha},\alpha_{1}} \leftarrow v$\\
  $t, i \leftarrow 0$\\
    $\alpha_{1} \leftarrow s$\\
\REPEAT
 \REPEAT
  \STATE $C^{i+1}_{N_{\alpha},\alpha_{t+1}} \leftarrow v +\alpha_{t+1} C^{i}_{N_{\alpha},\alpha_{t+1}} A$
   \STATE $i \leftarrow i + 1$
  \UNTIL{$C^{i}_{N_{\alpha},\alpha_{t}} - C^{i-1}_{N_{\alpha},\alpha_{t}} \le \epsilon $ or $i > k$}\\
   $t \leftarrow t+1$\\
   $C_{N_{\alpha},\alpha_{t}} \leftarrow  \frac{1}{ \sum_{j}^n {C}^{i}_{N_{\alpha},\alpha_{t}}(j)}C^{i}_{N_{\alpha},\alpha_{t}}$\\
   $\alpha_{t+1} \leftarrow \alpha_{t}+s$\\
   $i \leftarrow 0$\\
   $C^{0}_{N_{\alpha},\alpha_{t+1}} \leftarrow v$\\
\UNTIL{ $C_{N_{\alpha}, \alpha_{t}} = C_{N_{\alpha}, \alpha_{t-1}} = C_{N_{\alpha}}$ or $\alpha_{t}\ge1$}
\end{algorithmic}
\end{algorithm}

Algorithm~\ref{alg:1} does not depend on the value of $\lambda_1$.
Since normalized $\alpha$-centrality varies with $\alpha$ only for $\alpha < \frac{1}{|\lambda_1|}$, in order to study this variation, we may choose a step size of $s=\frac{c}{\min(d^{out}_{max},d^{in}_{max})}$, where $c<1$ is a constant. This is because, using the Gershgorin circle theorem, we know that $|\lambda_1| \le \min(d^{out}_{max},d^{in}_{max})$.

As can be seen in algorithm \ref{alg:1}, in each iteration, for a given value of $\alpha$, $C^{i+1}_{N_{\alpha},\alpha}$  depends only on $C^{i}_{N_{\alpha},\alpha}$ and $A$. Considering the network comprises of $n$ actors and $m$ links between them, in a naive implementation of algorithm \ref{alg:1}, each iteration has a runtime complexity of $O(m)$ and space complexity of $O(m+n)$. Assuming that the main memory just large enough to hold both $C^{i+1}_{N_{\alpha},\alpha}$ and  $C^{i}_{N_{\alpha},\alpha}$, the i/o cost for each iteration is $O(m)$.  If main memory is large enough to hold only  $C^{i+1}_{N_{\alpha},\alpha}$ , and assuming efficient data structure such as a sorted link list is used to store $A$, i/o cost is $O(m+n)$.
Since the formulation of normalized $\alpha$-centrality is very similar to that of PageRank (Equations \ref{a-cen} and \ref{eq:ppr}), similar block based strategies can be used for fast and efficient computation of both PageRank and normalized $\alpha$-centrality \cite{Haveliwala:1999} \cite{Kamvar:2003}. Like PageRank, normalized $\alpha$-centrality can easily be implemented using the map-reduce paradigm \cite{Dean:2008}, guaranteeing the scalability of this algorithm and its applicability to very large datasets.
Apart from normalized $\alpha$-centrality and PageRank,  we calculate the influence scores based on closeness centrality, graph centrality, betweenness centrality, in-degree centrality, out-degree centrality and SenderRank. Analogous to $\alpha$-centrality, for the other parametric measures of centrality, namely PageRank and SenderRank, we investigate the change in ranking as the value of the parameter changes.
Since this friendship network shows the small world phenomena (section \ref{collection}) and is unweighted, a fast approximation of betweenness centrality can be done with $O(km)$ run-time complexity where $k=\Theta(\frac{log \mbox{ } n}{\epsilon ^{2}})$, for $\epsilon>0$ \cite{Eppstein:2004}.
However, we use the fast algorithm for betweenness centrality given by Brandes \cite{brandes:2001}  for the calculation of betweenness centrality.  It has $O(n+m)$ space and $O(mn)$ run-time complexity.  Graph centrality and closeness centrality can be computed in very similar lines. 

An investigation into the stability of centrality measures when networks  are sampled was carried out in \cite{Costenbader:2003}. Eigen-vector centrality turned out to be the most robust centrality followed by in-degree centrality. For symmetric cases as $\alpha \to \frac{1}{|\lambda^{-}_{1}|}$, the eigen-vector centrality rank and $\alpha$-centrality ranks are identical  when $\lambda_1$ is strictly greater than any other eigenvalue \cite{Bonacich:2001}. In this paper, we prove that for normalized $\alpha$-centrality, $\lim_{\alpha \to \frac{1}{|\lambda_{1}|}} C_{N_{\alpha}, \alpha}$ exists and is equal to $C_{N_{\alpha}}$ which is independent of $\alpha$. We also show that for symmetric matrices, the rankings given by eigenvector centrality $C_{E}$ are equivalent to the rankings given by normalized $\alpha$-centrality $C_{N_{\alpha}}= \lim_{\alpha \to \frac{1}{|\lambda_{1}|}} C_{N_{\alpha}, \alpha}=C_{N_{\alpha}, \alpha >\frac{1}{\left |\lambda_{1}\right | }} $ (given that $\lambda_1$ is strictly greater than any other eigenvalue).

\subsection{Evaluation of Influence Predictions}
Next, we compare the predicted rankings using influence models described in Section~\ref{sec:influence} with the rankings obtained from the empirical estimate of influence using Pearson's correlation coefficient, since ties in rank exist.

 \begin{figure}[tbh]
\begin{tabular}{c}
  \includegraphics[height=1.8 in, width= 3.2 in]{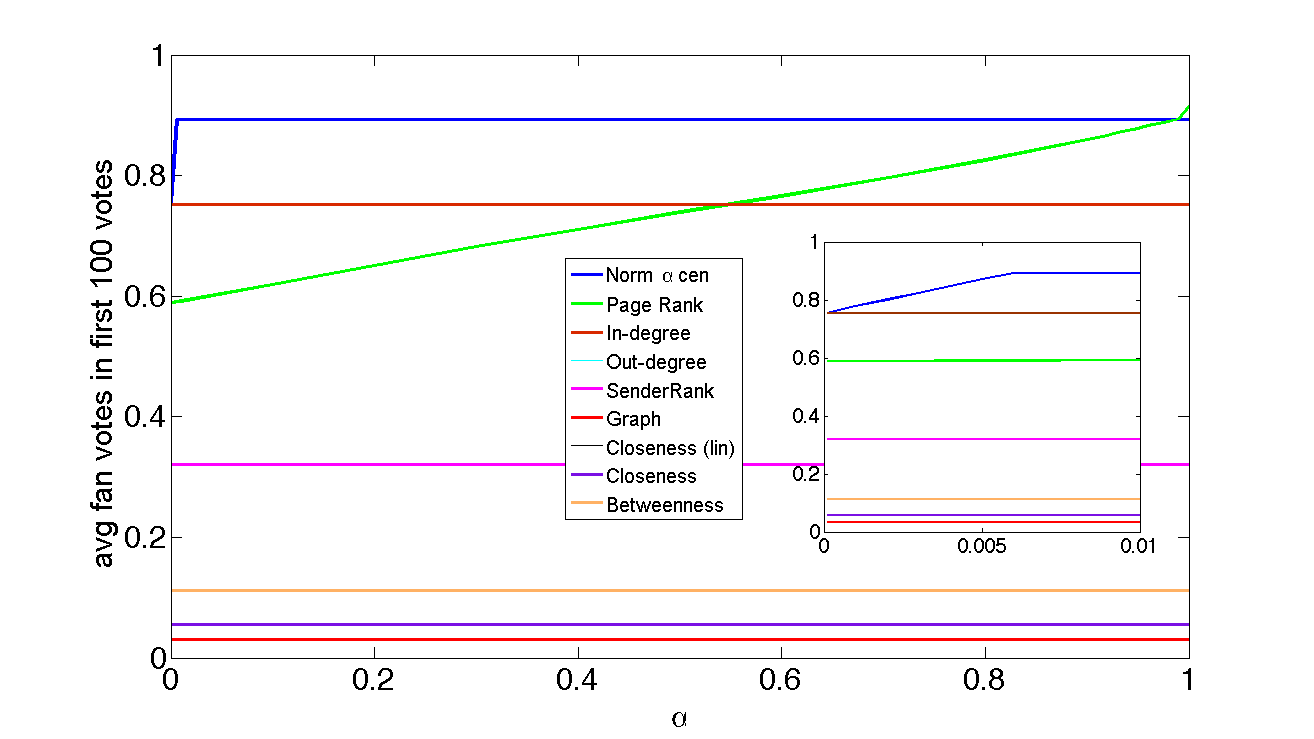}\\
  (a)\\
   \includegraphics[height=1.8 in, width=3.2 in]{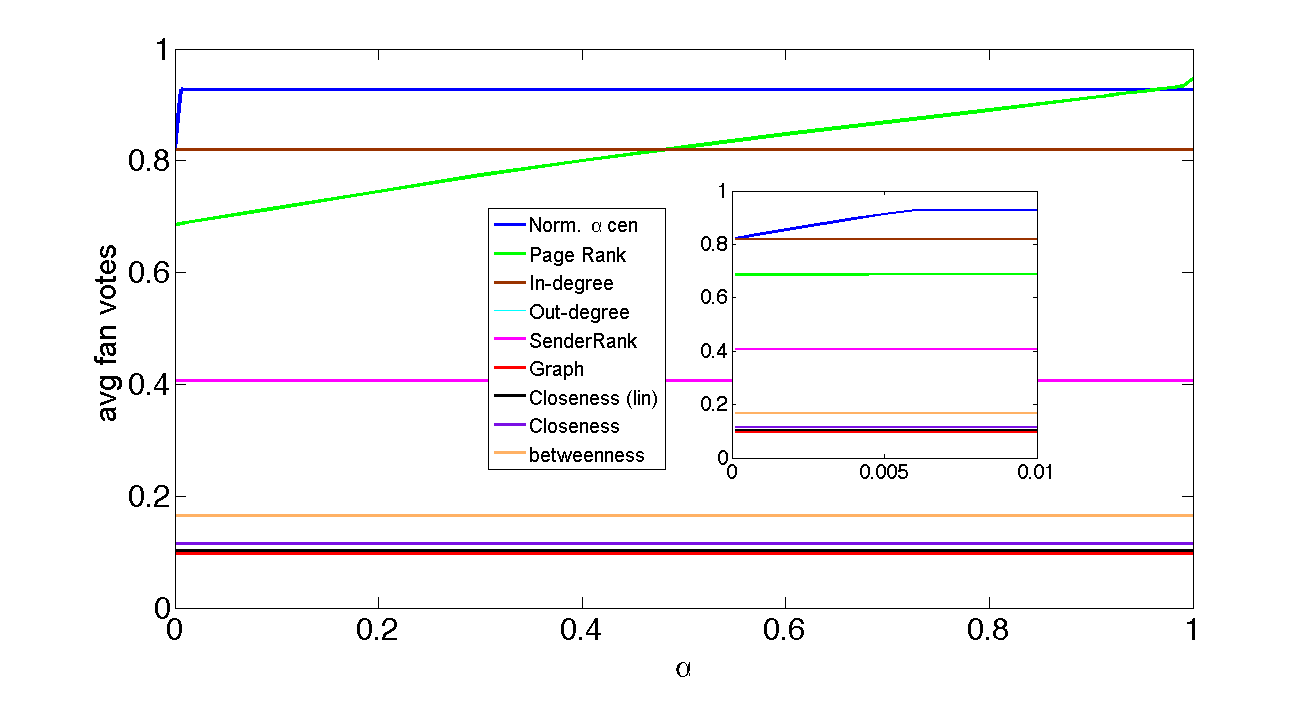}  \\
   (b)\\
\end{tabular}
\caption{ (a) Correlation between the rankings produced by the empirical measure of influence, which computes the number of fan votes within the first 100 votes, and rankings computed by different centrality measures.
(b) Correlation between the rankings produced by the empirical measure of influence, which computes the number of fan votes within all votes, and rankings computed by different centrality measures.
Note that $\alpha (0\le \alpha \le 1)$ stands for the \emph{attenuation factor}  for normalized $\alpha$-centrality and \emph{damping factor} for PageRank and SenderRank.  The inset zooms into the variation in correlation for $0 \le \alpha \le 0.01$  }\label{fig:correl_inset}
\end{figure}

Figure \ref{fig:correl_inset}(a)  shows correlation of influence rankings of the submitters, estimated from the average fan votes within the first 100 votes with their  influence rankings  relative to each other, calculated using different centrality measures. As can be seen from Equation  \ref{eq:pr}, \ref{a-cen} and \ref{eq:sr},  the characterization  of  \emph{damping factor}  (or restart probability)  in PageRank and SenderRank and the \emph{attenuation factor} in (normalized) $\alpha$-centrality are very similar mathematically. Therefore, without loss of generality, we represent both as $\alpha (0 \le \alpha \le 1 )$.

In  Section~\ref{Appendix} we prove that normalized $\alpha$-centrality $C_{N_{\alpha},\alpha}$ converges to $C_{N_{\alpha}} \forall \alpha \in ( \frac{1}{|\lambda_{1}|},1]$ ( $C_{N_{\alpha}, \alpha > \frac{1}{|\lambda_{1}|}}(i) =C_{N_{\alpha}}(i)$).  In Fig.~\ref{fig:correl_inset} (inset) we can clearly observe  this.  $C_{N_{\alpha},\alpha}$ (shown in blue) converges  to  $C_{N_{\alpha}}$ for $\alpha >0.007$. The correlation of $C_{N_{\alpha}}$ with the empirical estimate of influence is very high (corr. = 0.893024).

On the other hand, we observe that though the PageRank score converges for every value of $\alpha$, the PageRank score,  and hence the ranking, is dependent on the value of $\alpha$. Various studies have tested different damping factors, but it is generally assumed that the damping factor should be set around $\alpha=0.85$ \cite{Brin:1998}. Boldi et al. \cite{Boldi:2005} claim that in case of PageRank, ``for real-world graphs values of $\alpha$ close to 1 do not give a more meaningful ranking.''  Except for values $\alpha$ close to 1, the influence rankings calculated from normalized $\alpha$-centrality correlated better with the empirical estimates of influence rankings than PageRank rankings.

Correlation of SenderRank, $C_{SR}$ is 0.321. We also observe that in-degree centrality, $C_{d^{in}}$ is better correlated to the empirical estimate of influence (corr.= 0.753) than out-degree centrality, $C_{d^{out}}$ ( corr. = 0.32). Higher in-degree implies greater number of fans. A bigger, connected network of fans, fans of fans an so on can be inferred from higher (normalized) $\alpha$ centrality. Since the spread of information depends of number of users, who can see and spread the story a user submitted; fans and networks of fans  have a greater contribution to this spread than friends and network of friends. Hence (normalized) $\alpha$ centrality and in-degree centrality are better models for predicting influence than SenderRank and out-degree, when the underlying dynamic process is information propagation. There is also a high correlation between in-degree centrality and normalized $\alpha$-centrality. For  $C_{N_{\alpha}}$, the correlation is 0.91. Our results are in agreement with  a comparison study of centralities for biological networks \cite{Koschuetzki:2004}, where it was shown that  the correlation between  eigen-vector centrality and in-degree centrality was high.

Since \emph{non-conservative} flow of information on Digg is very different from the \emph{conservative} flow underlying geodesic path-based ranking measures, these measures are not well correlated to the empirical estimate of influence. Correlation of closeness centrality, $C_c$ \cite{Hakimi:1965},\cite{sabidussi:1966},\cite{Wasserman:1994} is 0.0564 and of that  due Lin et al. \cite{Lin:1976} is 0.0555. Correlation of graph centrality, $C_{g}$ is 0.0313 and of betweenness centrality $C_{b}$ is 0.1112. 

If we estimate the influence rankings of users, by taking the average number of fan votes in all the votes that their stories receive, the trends (Fig.~\ref{fig:correl_inset} (b) ) are very similar to those observed above (Fig. ~\ref{fig:correl_inset} (a)). All centrality measures are better correlated to the empirical estimate of influence thus obtained, as can be seen from Fig. \ref{fig:correl_inset}. The correlation of $C_{N_{\alpha}}$ with the empirical estimate of influence is very high (corr. = 0.928). Again the correlation of PageRank changes with $\alpha$ and except for $\alpha$ very close to 1, is less than that of (normalized) $\alpha$-centrality. As in Fig.~\ref{fig:correl_inset} (a), in-degree centrality,  $C_{d^{in}}$, is better correlated (corr.=0.82)  than out-degree centrality, $C_{d^{in}}$ (corr.=0.41). The correlation between $C_{d^{in}}$ and  $C_{N_{\alpha}}$ is 0.92. Correlation of closeness centrality, $C_c$ \cite{Hakimi:1965},\cite{sabidussi:1966},\cite{Wasserman:1994} is 0.116 and of that  due Lin et al. \cite{Lin:1976} is 0.103. Correlation of graph centrality, $C_{g}$ is 0.097 and of betweenness centrality $C_{b}$ is 0.1657. Correlation of SenderRank $C_{SR}$, is 0.407.

Next, we predict the rankings  of all 69,524 users within the network.  We do this using the influence models described above. Interestingly, the top user predicted by most models is `inactive'. `Inactive' is the nomenclature used by Digg to denote users who are no longer active, i.e., posting or voting on new stories. The connection between an `active'  user $u$  and another user $i$ exists even after $i$ has become inactive. Thus `inactive' user acts as a sink for these dangling links. We analyze the overall rankings of the  top 100 of the 289 submitters whose influence we have estimated empirically (using average fan votes in the first 100 votes). Let $emp$ be the set of rankings corresponding to the top 100 of 289 submitters as determined by the empirical influence. Let $pred$ be the set of corresponding rankings for the same submitters  using an influence model.
The probability that  the top 100 of these 289 submitters ($emp(i) \in [1,100]$) are among the top 100 of the  69,524 active users as predicted by model $pred$ is given by recall, $R= |emp\cap pred|/|emp|$.
Recall for normalized $\alpha$-centrality, $C_{N_{\alpha}}$, is high ($0.76$). Using in-degree centrality, $C_{d_{in}}$ for predictions,  reduces  recall to $0.6$. For PageRank $C_{pr,0.9}$  and  betweenness centrality $C_{b}$, recall is $0.29$ and $0.21$ respectively.  Recall is negligible when $C_{c}$,$C_{g}$, $C_{d_{out}}$ and $C_{SR}$ are used for prediction.

The results corroborate our hypothesis that,  since the underlying  non-conservative dynamic process of (normalized) $\alpha$-centrality, most closely resembles the dynamic process of information propagation in Digg, (normalized) $\alpha$-centrality is a better predictor of the influential users on Digg, than other influence models.

\section{Conclusion}
In this paper we emphasize the need to distinguish between different dynamic processes occurring in complex networks based on their distinct characteristics. Specifically, we categorize such processes into  {conservative} and  {non-con\-ser\-vative}, based on the nature of the flow. Further, we classify structural models which predict the influence standings of actors within a network into  {conservative} and  {non-conservative} models based on the underlying dynamic process that these models emulate. We stress that to get the best predictions of influence within an network using a influence model, the implicit underlying dynamic process of the model should have a close correspondence to the actual dynamic process taking place in that network.

Online social networks, provide us with a unique opportunity to study the continuously evolving dynamic processes within these networks. Here, we analyze \emph{information flow} on the social news aggregator Digg. We hypothesize that such a process is  {non-conservative} in nature. Hence to best predict the influential people within this network, we need a  {non-conservative} influence model. The ability to observe the actual dynamic process occurring on Digg, allows us to get an empirical estimate of influence within it. We prove that this estimate of influence is statistically significant. Using this empirical influence measure enables us to evaluate the predictions of different influence models.  To the best of our knowledge, this is the first work which evaluates influence models based on the structural properties of complex networks using the actual underlying dynamics of the network.

As hypothesized,  {non-conservative} models seem to perform better than  {conservative} models of influence. Specifically, we observed that the  {non-conservative} model of \emph{(normalized) $\alpha$-centrality} is the best predictor of influence within Digg, where the underlying dynamic process is information propagation. In this paper, we have also given a simple algorithm for computing normalized $\alpha$-centrality and the analytical proofs associated with it.

Future work would include applying similar analytical tools to  predict influentials on other online social networks. Most of the existing structural models of influence, assume that the structure of the network remains static in the course of study. However, online social networks are continually evolving. But researchers have not yet completely understood, the true nature of evolution of these networks. In future, we would like to delve deeper into the study of evolution of these networks; and apply the knowledge thus gained to build upon the existing prediction tools, to take into account the continual evolution of these networks.

\small
\bibliographystyle{abbrv}
\bibliography{sigproc}  
\balancecolumns
\section{Appendix}
\label{Appendix}

If $\lambda$ is an \emph{eigenvalue} of $A$, then
\begin{eqnarray}
(I-\frac{1}{\lambda}A)x=0
\end{eqnarray}
Invertibility of $(I-\frac{1}{\lambda}A)$ would lead to the trivial solution of eigenvector $x$( $x=0$).
Hence for computation of eigenvalues and eigenvectors, we require that no inverse of $(I-\frac{1}{\lambda}A)$  should exist, i.e.
\begin{equation}
Det(I - \frac{1}{\lambda} A)= 0
\label{eq:char_eq}
\end{equation}
Equation \ref{eq:char_eq} is called the \emph{characteristic equation} solving which gives the \emph{eigenvalues} and \emph{eigenvectors} of adjacency matrix $A$.

Using eigenvalues and eigenvectors, the adjacency matrix $A$ can be written as:
\begin{equation}
A=X\Lambda X^{-1} = \sum_{i=1}^{n} \lambda_{i}Y_{i}
\label{eq:eigen}
\end{equation}
where $X$ is a matrix whose columns are the eigenvectors of $A$.
$\Lambda$ is a diagonal matrix, whose diagonal elements are the eigenvalues, $\Lambda_{ii}=\lambda_{i}$, arranged according to the ordering of the eigenvectors in $X$.
Without loss of generality we assume that $\lambda_{1}> \lambda_{2} >\cdots >\lambda_{n}$.
The matrices $Y_{i}$ can be determined from the product
\begin{equation}
Y_{i}=X {Z}_{i}X^{-1}
\label{eq:z}
\end{equation}
where $Z_{i}$ is the \emph{selection matrix} having zeros everywhere except for element ${(Z_i)}_{ii}=1$ ~\cite{Gebali:2008}.

The \emph{$\alpha$-centrality matrix} $C_{\alpha,k}$  $\forall \alpha \in [0,1]$is given by:
\begin{eqnarray}
C_{\alpha,k} &=& I+ \alpha A+\alpha^2 A^2+\cdots+\alpha^{k} A^{k} \nonumber  \\
&=&{\displaystyle \sum_{t=0}^k} \alpha^{t}A^{t}
\label{eq:bc_eq}
\end{eqnarray}

The \emph{normalized  $\alpha$-centrality matrix} is then given by:
\begin{equation}
NC_{\alpha,k}=\frac{1}{\displaystyle \sum_{i,j}^{n} {(C_{\alpha,k})}_{ij}}C_{\alpha,k}
\end{equation}
As shown in Equation \ref{eq:b-cen}  and \ref {eq:norm-cen} $\alpha$-centrality vector is $vC_{\alpha, k \to \infty}$ and normalized $\alpha$-centrality vector is $vNC_{\alpha, k \to \infty}$.

$A^{k}$ can then be written as :
\begin{equation}
A^{k}=X{\Lambda}^{k} X^{-1} =\sum_{i=1}^{n} {\lambda_{i}^{k}}Y_{i}
\label{eq: lambda_k}
\end{equation}
Using Equation \ref{eq: lambda_k}, \ref{eq:bc_eq} reduces to
\begin{eqnarray}
C_{\alpha,k}&=&{\displaystyle \sum_{i=1}^n}{\displaystyle \sum_{t=0}^k} \alpha^{t}{\lambda_{i}^{t}}Y_{i}\nonumber \\
&=&{\displaystyle \sum_{i=1}^n} \frac{{(-1)}^{p_i}(1-\alpha^{k+1} \lambda_{i}^{k+1})}{{(-1)}^{p_i}(1-\alpha \lambda_{i})} Y_{i} \nonumber \\
\label{eq:cm}
\end{eqnarray}
where $p_{i}=0$ if $\alpha \left |\lambda_{i}\right | <1$ and $p_{i}=1$ if $\alpha \left |\lambda_{i}\right | >1$.
As obvious from above, for  equation \ref{eq:bc_eq} and \ref{eq:cm}  to hold non-trivially,  $\alpha \neq \frac{1}{\left |\lambda_{i}\right|} \forall i \in 1,2\cdots,n$.

We consider the characterization of the series \{ $NC_{\alpha, k \to \infty}$ \}for $\alpha \in [0,1]$.
 \begin{enumerate}
\item  $\alpha\ll\frac{1}{\left| \lambda_{1}  \right|}$: If $\alpha\ll\frac{1}{\left| \lambda_{1}  \right|}$ , $C_{\alpha, k \to \infty}$ (and $NC_{\alpha, k \to \infty}$ ) would be independent of $\alpha$, since
\begin{eqnarray}
C_{\alpha, k \to \infty} \approx I \nonumber \\
NC_{\alpha, k \to \infty} \approx \frac{1}{n}I
\label{eq:b_alpha1}
\end{eqnarray}
\item   $\alpha<\frac{1}{\left |\lambda_{1}\right | }$: The sequence of matrices $\{ C_{\alpha,k} \}$ would \emph{converge} to $C_{\alpha} $ as $k \to \infty $ if all the sequences $ \{{(C_{\alpha , k})}_{i j}\}$ for every  fixed $i$ and $j$ converge to ${(C_{\alpha})}_{i j}$ ~\cite{Dienes:1932}.
If  $\alpha<\frac{1}{\left |\lambda_{1}\right | }$,  $ {C_{\alpha,k}}$ \emph{converges} to $C_{\alpha}$.
\begin{eqnarray}
C_{\alpha} &= &C_{\alpha,k \to \infty}={\displaystyle \sum_{i=0}^n} \frac{1}{1-\alpha \lambda_{i}} Y_{i} = {(I- \alpha A)^{-1}} \nonumber \\
&  &NC_{\alpha, k \to \infty} = \frac{C_{\alpha}}{\sum^{n}_{ij} {(C_{\alpha})}_{ij}}
\label{eq:b_alpha}
\end{eqnarray}
\item  $\alpha$> $\frac{1}{\left |\lambda_{1}\right | } $ and $k \to \infty$, ${\alpha^{k}A^{k}}$  dominates in the Equation \ref{eq:cm}.
\begin{eqnarray}
C_{\alpha,k\to \infty} \approx {\alpha^{k}A^{k}} \nonumber \\
NC_{\alpha,k\to \infty} \approx \frac{1}{\displaystyle \sum_{i,j}^{n} {A}^{k}_{ij}}A^{k}
\label{eq:maxA}
\end{eqnarray}
\end{enumerate}

\begin{theorem}
\label{th1}
The induced ordering of nodes due to normalized $\alpha$-centrality would be equal to the induced ordering of nodes due to $\alpha$-centrality for $\alpha <\frac{1}{\left |\lambda_{1}\right | }$.
\end{theorem}
\begin{proof}
Since $C_{alpha, \alpha}= vC_{\alpha, k \to \infty}$ and $C_{N_{\alpha}, \alpha}= vNC_{\alpha, k \to \infty}$, from equations \ref{eq:b_alpha1} and \ref{eq:b_alpha}, the  induced ordering of nodes due to $\alpha$-centrality ($\alpha <\frac{1}{\left |\lambda_{1}\right | }$) would be equal to  induced ordering of nodes due to normalized $\alpha$-centrality  ($\alpha <\frac{1}{\left |\lambda_{1}\right | }$).
\end{proof}

\begin{theorem}
\label{th2}
The value of normalized $\alpha$-centrality remains the same $\forall \alpha \in (\frac{1}{\left |\lambda_{1}\right | },1]$ ( $C_{N_{\alpha}, \alpha >\frac{1}{\left |\lambda_{1}\right | }  } = C_{N_{\alpha}}$).
\end{theorem}
\begin{proof}
As can be seen from equation \ref{eq:maxA}  when $\alpha$> $\frac{1}{\left |\lambda_{1}\right | } $ and $k \to \infty$, $NC_{\alpha,k \to \infty}$ reduces to  $\frac{1}{\displaystyle \sum_{i,j}^{n} {A}^{k}_{ij}}A^{k}$ and is independent of $\alpha$. Since normalized $\alpha$-centrality,  $C_{N_{\alpha}, \alpha} =v NC_{\alpha, k \to \infty}$, therefore, value of normalized $\alpha$-centrality value remains the same $\forall \alpha \in (\frac{1}{\left |\lambda_{1}\right | },1]$ ($C_{N_{\alpha}, \alpha >\frac{1}{\left |\lambda_{1}\right | }  } = C_{N_{\alpha}}$).
\end{proof}

 The remaining theorems hold under the condition that $|\lambda_{1}|$ is strictly greater than any other eigenvalue, which is true in most real life cases studied.
\begin{theorem}
\label{th3}
$\lim_{\alpha  \to \frac{1}{|\lambda_{1}|}}C_{N_{\alpha}, \alpha}$ exists and $\lim_{\alpha  \to \frac{1}{|\lambda_{1}|}}C_{N_{\alpha}, \alpha} =C_{N_{\alpha}}=C_{N_{\alpha}, \alpha >\frac{1}{\left |\lambda_{1}\right | }  } =\frac{vY_{1}}{ \sum_{i,j}^{n} {(Y_{1})}_{ij}} $.
\end{theorem}
\begin{proof}
Under the assumption that $|\lambda_{1}|$ is strictly greater than any eigenvalue,  Equation  \ref{eq:b_alpha}, as $\alpha \to \frac{1}{|\lambda^{-}_{1}|}$ reduces to
\begin{equation}
C_{\alpha \to \frac{1}{|\lambda^{-}_{1}| }, k \to \infty}=  \frac{1}{1-\alpha \lambda_{1}}Y_{1}
\end{equation}
This is because all other eigenvectors shrink in importance as $\alpha \to \frac{1}{|\lambda^{-}_{1}|}$ \cite{Bonacich:2001}. Therefore as $\alpha \to \frac{1}{|\lambda^{-}_{1}|}$ , we have
\begin{equation}
NC_{\alpha  \to \frac{1}{|\lambda^{-}_{1}|}, k \to \infty} = \frac{1}{\displaystyle \sum_{i,j}^{n} {(Y_{1})}_{ij}}Y_{1}
\label{eq:max1}
\end{equation}
Under the assumption that $|\lambda_{1}|$ is strictly greater than any other eigenvalue, ${\alpha^{k}\lambda_{1}^{k}Y_{1}}$  dominates in the Equation \ref{eq:cm},  \ref{eq:maxA}.
\begin{eqnarray}
C_{\alpha,k\to \infty} \approx {\alpha^{k}\lambda_{1}^{k}Y_{1}} \nonumber \\
NC_{\alpha,k\to \infty} \approx \frac{1}{\displaystyle \sum_{i,j}^{n} {(Y_{1})}_{ij}}Y_{1}
\label{eq:max}
\end{eqnarray}

Hence from equation \ref{eq:max},  as $\alpha \to \frac{1}{|\lambda^{+}_{1}|}$,  we have
\begin{equation}
NC_{\alpha  \to \frac{1}{|\lambda^{+}_{1}|}, k \to \infty} = \frac{1}{\displaystyle \sum_{i,j}^{n} {(Y_{1})}_{ij}}Y_{1}
\label{eq:max2}
\end{equation}

Since $\lim_{\alpha  \to \frac{1}{|\lambda^{-}_{1}|}}NC_{\alpha, k \to \infty}  = \lim_{\alpha  \to \frac{1}{|\lambda^{+}_{1}|}}NC_{\alpha, k \to \infty} =\frac{Y_{1}}{\sum_{i,j}^{n} {(Y_{1})}_{ij}}$, therefore
$\lim_{\alpha  \to \frac{1}{|\lambda_{1}|}}NC_{\alpha, k \to \infty}$ exists and
\begin{equation}
\lim_{\alpha  \to \frac{1}{|\lambda_{1}|}}NC_{\alpha, k \to \infty}= \frac{Y_{1}}{ \sum_{i,j}^{n} {(Y_{1})}_{ij}}.
\label{eq:max3}
\end{equation}
Since $C_{N_{\alpha}, \alpha}= vNC_{\alpha, k \to \infty}$, therefore,  $\lim_{\alpha  \to \frac{1}{|\lambda_{1}|}}C_{N_{\alpha}, \alpha} =C_{N_{\alpha}}=C_{N_{\alpha}, \alpha >\frac{1}{\left |\lambda_{1}\right | }  } =\frac{vY_{1}}{ \sum_{i,j}^{n} {(Y_{1})}_{ij}} $.

\end{proof}

\begin{theorem}
\label{th4}
For symmetric matrices, the  induced ordering of nodes due to eigenvector centrality $C_{E}$ is equivalent to the  induced ordering of nodes given by normalized centrality $C_{N_{\alpha}}= \lim_{\alpha  \to \frac{1}{|\lambda_{1}|}}C_{N_{\alpha}, \alpha}=C_{N_{\alpha}, \alpha >\frac{1}{\left |\lambda_{1}\right | }}=\frac{vY_{1}}{ \sum_{i,j}^{n} {(Y_{1})}_{ij}} $.
\end{theorem}
\begin{proof}
For symmetric matrices
\begin{equation}
A=X\Lambda X^{-1} = X \Lambda X^{T}
\label{eq:eigen1}
\end{equation}
Therefore equation \ref{eq:z} reduces to
\begin{equation}
Y_{i}=X {Z}_{i}X^{T} = X_{i}X_{i}^{T}
\end{equation}
\noindent where $X_{i}$ is the column of $X$ representing the eigenvector corresponding to $\lambda_{i}$.
Hence, in case of symmetric matrices:
\begin{eqnarray}
C_{N_{\alpha}} &= & C_{N_{\alpha}, \alpha >\frac{1}{\left |\lambda_{1}\right | }} = \lim_{\alpha  \to \frac{1}{|\lambda_{1}|}}C_{N_{\alpha}, \alpha} \nonumber \\
&=& \frac{vY_{1}}{ \sum_{i,j}^{n} {(Y_{1})}_{ij}}  \nonumber \\
&=& c_{1}vX_{1}X_{1}^{T} = c_{2}X_{1}^{T}
\label{eq:max4}
\end{eqnarray}
where $c_{1}= \frac{1}{\sum_{i,j}^{n} {(Y_{1})}_{ij}}$ and $c_{2}= c_{1}vX_{1}$.

Since $X_{1}^{T}$ corresponds to the eigenvector centrality vector $C_{E}$, hence for symmetric matrices, the induced ordering of nodes given by eigenvector centrality $C_{E}$ is equivalent to the  induced ordering of nodes given by normalized centrality $C_{N_{\alpha}}=  \lim_{\alpha  \to \frac{1}{|\lambda_{1}|}}C_{N_{\alpha}, \alpha}=C_{N_{\alpha}, \alpha >\frac{1}{\left |\lambda_{1}\right | }}=\frac{vY_{1}}{ \sum_{i,j}^{n} {(Y_{1})}_{ij}} $.
\end{proof}

\end{document}